# Nano-confinement induced nucleation of ice-VII at room temperature

Jonggeun Hwang[1], Dongha Shin[1,2], Brendan T. Deveney[3], Manhee Lee[3,4,*], Xingcai Zhang[3,5*] & Wonho Jhe[1,3*]

[1]Center for 0D Nanofluidics, Institute of Applied Physics, Department of Physics and Astronomy, Seoul National University, Seoul 08826, Korea

[2]Department of Chemistry and Chemical Engineering, Inha University, Incheon 22212, Korea

[3]John A. Paulson School of Engineering and Applied Sciences, Harvard University, Cambridge, MA 02138, USA

[4]Department of Physics, Chungbuk National University, Cheongju, Chungbuk 28644, Korea

[5]School of Engineering, Massachusetts Institute of Technology, Cambridge, MA 02139, USA

*Corresponding author. E-mail: whjhe@snu.ac.kr, mlee@cbnu.ac.kr, xingcai@seas.harvard.edu

**The hydrogen bond (HB) network of water under confinement has been predicted to have distinct structures from that of bulk water. However, direct measurement of the structure has not been achieved. Here, we present experimental evidence of confinement-induced ice formation in water. We directly probe the HB network of a water nano-meniscus formed and confined between a mica substrate and a precisely-controlled-plasmonically active silver tip. By employing tip-enhanced Raman spectroscopy (TERS), we observe a novel double donor-double acceptor (DDAA) peak that emerges in the OH stretching band of water molecules at room temperature and at sub-nanometer confinement. This Raman peak indicates the presence of a solid phase of water, namely ice-VII with the body-centered cubic**

**(bcc) unit. Interestingly, we observe a structural transition from bcc DDAA (ice-VII) to tetrahedral DDAA as the confinement is weakened. Moreover, by identifying the spatial distribution of the HB network, we find that the bcc DDAA network of ice-VII is predominantly present within the interior of the confined water, rather than at air/water or at solid/water interfaces. This suggest the possibility that the appearance of ice-VII in the strongly confined space could be a general characteristic of water under extreme confinement.**

For low-dimensional water in confined geometries, its entropy and interfacial energy are modified due to the reduced degrees of freedom of its constituent molecules. The resulting changes in the physical, chemical and biological interactions with confined water are therefore associated with changes in the HB network of water[1–7]. Evidence of ordered structures in confined water have been reported in previous nano-confined water studies using atomic force microscopy (AFM)[8,9] and a surface force apparatus (SFA)[10]. The layered molecular structure of confined water and the increased mechanical relaxation time with respect to bulk water in these studies indicates that solid-like behaviors induced by confinement. However, conventional force-measurement methods cannot discern the structure of the HB network of water under confinement. In the most recent first-principle molecular dynamics (MD) simulations, water confined within nanoslits exhibits a various novel ice phase at room temperature, such as superionic phase[7] for monolayer water and bilayer ice-VII[11] for bilayer water, differing from the typical hexagonal ice structure. Despite this predication, quantitative experimental validation remains unsolved.

Despite its importance and interest, it has been extremely challenging to experimentally investigate the molecular structure of very small volumes of confined water. This is mainly attributed to the necessity for sub-nanometric control of confinement and the capability to resolve the spectroscopic profile of water, which has a small Raman cross-section and a large bandwidth in the OH stretching band. Various *in situ* spectroscopic experiments to study the molecular structure of confined water have been reported such as surface-enhanced Raman spectroscopy[6,12], sum-frequency generation spectroscopy[13], and X-ray spectroscopy[14]. However, spectroscopic techniques alone are insufficient to obtain the mechanical properties of confined water.

Here, we address the unexplored problem by quantitative and simultaneous measurements of Raman spectroscopy and force spectroscopy of confined water. To simultaneously obtain the molecular structural information and the mechanical properties, we build tip-enhanced Raman spectroscopy (TERS) setup which combines atomic force microscopy with confocal Raman spectroscopy. TERS is a near-field optical platform that offers the capability to probe confinement-dependent molecular structure of small quantities of molecules and provides significantly enhanced sensitivity compared traditional micro-Raman methods[15–17].

We study the structural and mechanical behavior of water confined between a sharp silver tip and a flat mica substrate as described in Figs. 1a and 1b. The mica substrate serves as an atomically flat and hydrophilic surface on which a meniscus can be spontaneously forms by capillary condensation when the tip-substrate distance $d$ within a few nanometers. To probe the mechanical properties of the condensed water, we employed a quartz tuning-fork (QTF) as a sensitive piezoelectric force sensor having a high quality factor ($\sim 10^4$) and a high stiffness ($\sim 10^4$ N/m), which allows precise control of confinement in terms of $d$. The QTF is operated in shear mode at its resonance ($\omega/2\pi \approx 32{,}745$ Hz) with an oscillation amplitude of 0.29 nm. Measurements are made at room temperature and at low relative humidity of approximately 1%. This environment creates a meniscus shape of small volume of water between the tip and the substrate, which is reliably measurable with our probe. (see Supplementary discussion S1 for AFM force curve at different RH). We measure the changes in the amplitude and phase induced by the confined water[18], to determine the elastic ($k_{int}$) and damping ($b_{int}$) interaction with water, as well as the corresponding mechanical relaxation time[9,19], $\tau_r = k_{int}/b_{int}\omega^2$, which serve as indicative parameters of solid-like behavior.

We calculate the electric-field distribution at the apex of the tip using Finite-difference time-domain (FDTD) simulations (COMSOL Multiphysics, Stockholm, Sweden). The enhanced local electric-field is generated near the sharp metal tip as a response of external field due to plasmonic resonance, resulting in high signal-to-noise ratio compared to traditional Raman spectroscopy. The silver tip used in our experiments provides an enhancement factor (EF) of over $2.5 \times 10^5$. The tip was fabricated by high-vacuum silver coating with an e-beam evaporator, demonstrating the high EF. Two different tip fabrication methods were also tested; low-vacuum silver coating with sputter and electrochemical etching (see Supplementary discussion S2 for other methods and their results). We also calculate the profile of the water meniscus between the tip and the mica substrate using the Young-Laplace equation[20], which describe the pressure difference across the non-planer liquid interface, at the same geometrical conditions with FDTD simulation. The calculated size of the plasmonic 'hotspot' is comparable to the size of water-meniscus, as illustrated in Fig. 1c. This indicates that the Raman spectra we acquire cover the entire confined water between tip and substrate (see Fig. S1 for details of the calculation results).

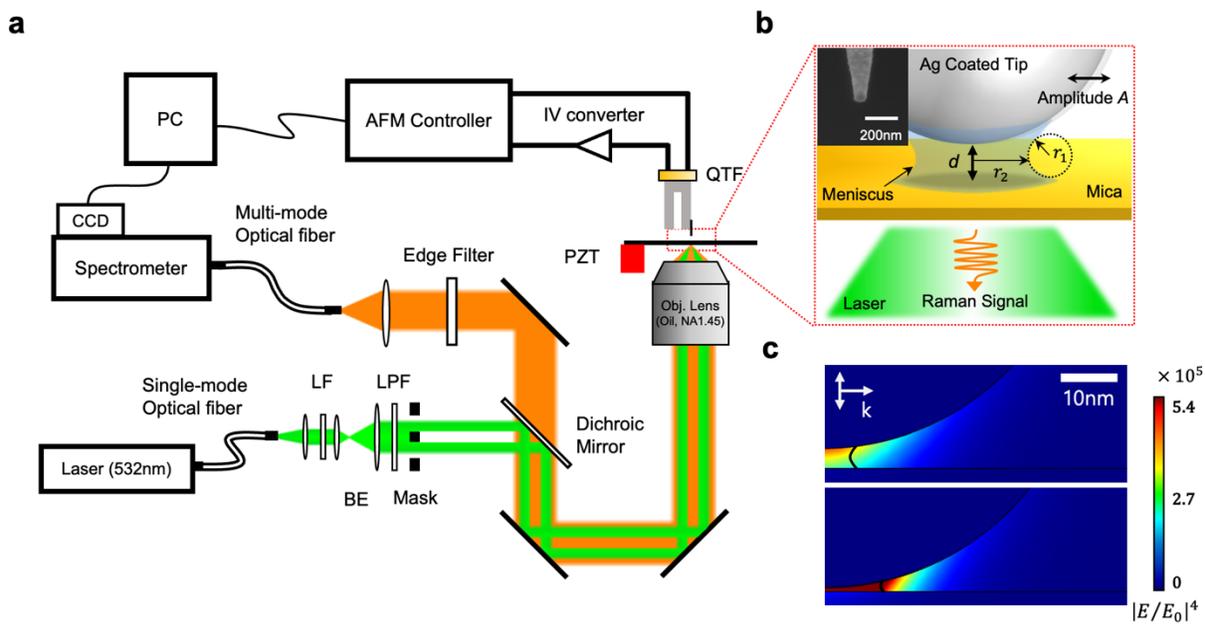

**Fig. 1. Experimental schematic of AFM-based TERS for molecular structure-force measurement. a**, TERS combined with QTF-based atomic force microscopy (AFM). BE; beam expander, LF; line filter for 532 nm laser, LPF; linear polarizer filter, PZT; piezoelectric transducer, objective lens with NA = 1.45. **b**, Schematic of a water nano-meniscus capillary-condensed in the tip-mica gap. The tip oscillates with amplitude $A$. $r_1$ and $r_2$ are the two principal radii-of-curvature of the meniscus. The inset shows the SEM image of a typical 80 nm-diameter silver-evaporated cantilever tip. **c**, FDTD simulation (color map) of EF for a given meniscus geometry (black curve) calculated by the Young-Laplace equation at $d$ = 3.0 nm (upper) and 0.5 nm (lower).

The Raman signal and the corresponding mechanical response of confined water are shown in Figs. 2a, 2b. We characterize three distinct regimes of $d$. The first is the non-contact regime [i]; no condensed water is formed in the nano-gap, resulting in constant amplitude and phase signals, and no Raman signal from confined water is observed. The second is the weak-confinement regime [ii]; the amplitude and the phase of tip change suddenly at $d \sim$ 3 nm due to the capillary condensation of water in the gap. In the Raman spectrum, there is an OH stretching band appears in the high-frequency region and there is no sign of signal in the low-frequency region. Note that water remains condensed between the tip and the substrate even when the tip is retracted to ~ 4.5 nm. The third regime is the strong-confinement regime within ~ 2 nm from the surface [iii]; as the tip approaches the substrate, there are considerable changes in the amplitude and phase signals and the corresponding Raman spectra even with a small change of $d$. Surprisingly, we observe the emergence of the lattice vibration bands in the low-frequency region and a peak at 3316 cm$^{-1}$ in the high-frequency region. The emergence of the lattice vibration band, which is commonly observed in many ice phases[21–23]. This indicates an ice-like HB network of water molecules in the strong-confinement regime. Since the presence of mica signals complicates the distinction between confined water and mica, we subtract mica signal from the raw data. Specifically, two peaks at 195 cm$^{-1}$ and 265 cm$^{-1}$ in the low-frequency region, and free-OH peak at 3610 cm$^{-1}$ in the high-frequency region are attributed to mica[24]. They are subtracted from the raw spectra for clarity, and resulting spectra in low-frequency region are fitted with 2 Gaussian curves as shown in Fig. 2a. The peak at 210 cm$^{-1}$ and the peak at 305 cm$^{-1}$ correspond to the DDAA and DA component of the OH stretching band, respectively[12]. In remaining spectra in high-frequency region, a CH$_n$ (n=2, 3) stretching band signal at 2900 cm$^{-1}$ from unwanted impurity molecules is observed, potentially originating from atmospheric carbon contaminant (ACC) or carbonaceous decomposition products[25–27] inhaled during the tip-fabrication process or during the chamber-purging process. Thus, we subtract the time-dependent CH$_n$ stretching band signals from each spectrum and decompose the resulting spectra into five Gaussian curves, as shown in Fig. 2a. (see Supplementary discussion S5 for full spectrum).

In the case of weakly-confined water, the measured spectrum is consistent with that of bulk water, but the specific peak positions and the absence of the lattice vibration mode in this regime suggest that weakly confined water is in a supercooled state. The OH stretching band of bulk water is decomposed into five different water types depending on the HB network type: DAA, DDAA, DA, DDA and free-OH, where D (A) stands for the HB donor (accepter). We can clearly identify the four different water types: DAA, DDAA, DA, and DDA in weakly confined water. Notice that the DDAA signal at 3178 cm$^{-1}$ (blue peak in the lower panel of Fig. 2c) is red-shifted and has a narrower bandwidth than bulk water. Therefore, the HB network of weakly-confined water is more likely correspond to the tetrahedral DDAA structure, characteristic of ice-I$_h$, rather than bulk liquid water. However, the DDAA signal is not accompanied by a lattice vibration signal as shown in the low-frequency region of Fig. 2a. This indicates the absence of an ordered HB network, suggesting that water in the weakly confined regime is closer to supercooled state than the solid ice-I$_h$. To quantify the reduced energy of the weakly-confined water in terms of temperature, we compare the measured DDAA position of the weakly-confined water with that of supercooled water[28], which yields approximately 224K (Fig. 2d). Note that the value does not represent the local temperature of confined water because our experiment is conducted at room temperature.

In the case of strongly-confined water, we find the emergence of the Raman peak at 3316 cm$^{-1}$, depicted by a red curve in the strongly-confined regime [iii] of Fig. 2a. This signal does not correspond to any known Raman components of bulk water. However, it exhibits a strong correlation with the lattice vibration signal that appears in the low-frequency band, commonly observed in the ice. Since the DDAA is commonly used as a structural indicator of solid phase of water[23,29], we compare the Raman peak at 3316 cm$^{-1}$ with the DDAA peak positions of various known ice phases (Fig. 2e) to identify its structure. Most ice structures exhibit DDAA peaks located below 3250 cm$^{-1}$; however, ice-VII at high pressure and at room temperature displays a peak near 3310 cm$^{-1}$ which closely aligns our new signal at 3316 cm$^{-1}$. Meanwhile, the observed

Raman signal is difficult to explain with other high-pressure ices such as ice-VIII and ice-X. Our experimental conditions are outside of the known stable region of bulk ice-VIII[30]. Moreover, we previously performed the surface-enhanced Raman spectroscopy (SERS) of confined water using silver nanoparticles (AgNPs), in which we observed the transition of ice-VII to ice-VIII at approximately 280 K. (See Fig. S2 for details of SERS experiment.) The ice-X is also excluded from candidate since the OH stretching band in ice-X is absent therein[31]. Consequently, we assign our new DDAA signal to the unit structure of ice-VII, i.e., bcc DDAA.

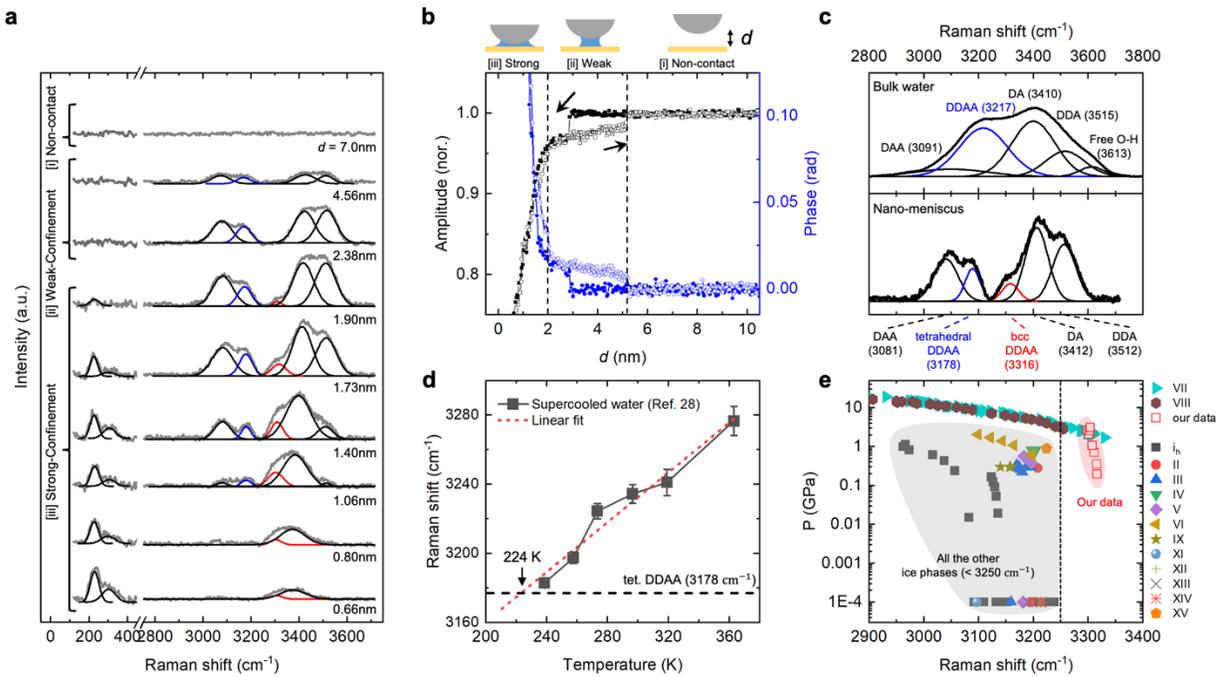

**Fig. 2. TERS spectra of single nano-meniscus and its spectral assignment. a**, The Raman spectra of the confined water are measured at each $d$. The lattice vibration modes are measured in low-frequency region, and the OH stretching bands are measured in the high-frequency region. **b**, Typical amplitude and phase response of QTF during tip approach (filled symbols) and retraction (open symbols). Determination of the zero-point ($d$ = 0) is described in Supplementary discussion S3. **c**, Comparison of the Raman spectrum taken at $d$ = 1.73 ± 0.09 nm and the Raman signal of bulk water taken at 300K. **d**. DDAA peak position of supercooled water[28] with respect to temperature. The dashed horizontal line is the tetrahedral DDAA (3178 cm$^{-1}$) in Fig. 2c. **e**, The DDAA peak position of various ice phases (ice-I$_h$, II, III, IV, V, VI, VII, VIII, IX, XI, XII, XIII, XIV, XV) versus pressure (see Supplementary discussion S4 for references therein). Our experimental data of bcc DDAA are also plotted (red open squares).

The HB networks including DAA, DDAA, DA, DDA, and free-OH are distributed homogeneously in bulk water[32] without interfaces. However, the confined water consists of two separate spatial parts: an interface and an interior. These regions can be occupied by different types of HB networks, as discussed in studies involving water trapped inside of carbon nanotubes (CNTs)[33], Ice clusters[34], and confined between Ag nanoparticles[12]. Both simulations[35] and experiments[12,33,34] on confined water confirm that the DDAA water is present in the interior, while other types of water are at the interface. Here, we further divide the interfacial region into the air/water (a/w) interfacial region with an interfacial thickness of $t_{a/w}$ and the solid/water (s/w) interfacial region with a thickness $t_{s/w}$. Therefore, the confined water is composed of three parts as shown in Fig. 3d-f; a/w region (green), s/w region (blue), and the interior region (red).

To determine where the bcc DDAA of ice-VII is formed in the confined water, we perform a spatial distribution analysis of HB networks. To correlate three spatial regions of confined water with different types of HB networks, we compare the normalized Raman intensity of the OH stretching band with the normalized EF-weighted volume and minimize their difference by optimizing two interfacial thickness $t_{a/w}$ and $t_{s/w}$. Here, the normalized EF-weighted volume is obtained by integrating the infinitesimal volume multiplied by the EF (Figs. 3a-c) over each partial region, a/w, s/w, or interior region (Figs. 3d-f). To determine optimal $t_{a/w}$ and $t_{s/w}$, we initially generate the EF-weighted volume maps for three spatial regions (a/w, s/w, and interior) at a given $t_{a/w}$ and $t_{s/w}$ and calculate three errors between the normalized EF-weighted volume map and the normalized Raman intensity for three regions at each $d$. Subsequently, we repeat these calculations for different $t_{a/w}$ and $t_{s/w}$ to obtain the total error map, given as the sum of the absolute values of the three errors, at each $d$ (Figs. 3g-i). The total error reaches the minimum as low as 1%, in the purple-colored area in Figs. 3g-i. (see Supplementary discussion S7 for details of the Spatial distribution analysis). The minimum points at each $d$ is plotted in Fig. 3k. Note that $t_{s/w}$ is independent of $d$ and is about 0.32nm, the monolayer thickness of water, in agreement with the results of the pre-adsorbed water layer on mica[8,36], whereas $t_{a/w}$ is 2 ~ 3 molecular thickness when $d > 1$ nm and of monolayer thickness when $d < 1$ nm. By comparing the EF-weighted volumes and the Raman intensity (Fig. 3j), we conclude that the a/w interfacial water consists of DDA and DAA, the s/w interfacial water contains DA, and the interior water is occupied by DDAA, with a tetrahedral structure in the case of weakly-confined water and a bcc structure in the case of strongly-confined water. The Spatial distribution analysis yields consistent results across experimentally accessible variations of the tip-water contact angles (see Fig. S3 and S4 for details).

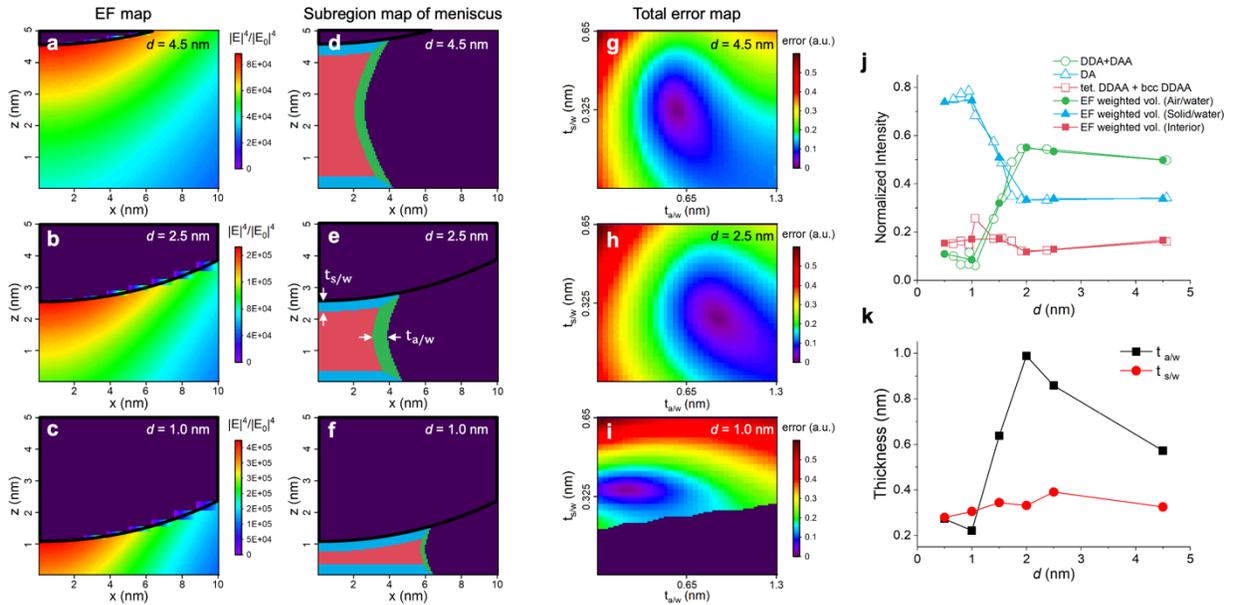

**Fig. 3. Spatial assignment of HB network. a-c,** EF map calculated using FDTD simulations for $d$ = 4.5 nm, $d$ = 2.5 nm, and $d$ = 1.0 nm, respectively. **d-f**, Meniscus profiles with area separation of three regions, where blue-colored regions represent the solid/water interfacial water, green-colored region represent the air/water interfacial water, and red-colored region represent the interior water. The black solid curve indicates the tip boundary. **g-i**, The total error map for given $d$. The optimized $t_{a/w}$ and $t_{s/w}$ are determined at which the error reaches a minimum, the violet-colored region of each error map. The error is not defined at the dark purple-colored area at the lower part of the **i** because the sum of the partial EF-weighted volume exceeds the total EF-weighted volume. **j**, the normalized Raman signal of selected combinations of peaks (open symbols) and the EF-weighted volume of nano-meniscus with optimized $t_{a/w}$ and $t_{s/w}$. (filled symbols) **k**, Optimized values of $t_{a/w}$ and $t_{s/w}$ are plotted versus $d$.

We observe evidence of a structural transition between tetrahedral DDAA and bcc DDAA as the confinement proceeds in strongly confined regime. Both the Raman shift (Fig. 4a) and the normalized Raman intensity (Fig. 4b) show the transition in molecular structure of water with changes in $d$, where the gray-colored transition region lies between $d$ of 1.06 ± 0.16 nm and 1.90 ± 0.12 nm. Moreover, the mechanical relaxation time $\tau_r$, the timescale of relaxation after deformation[9], increases as the confinement becomes stronger in strong-confinement (Figs. 4c, d). This suggest that the confined water is mechanically solid-like in the strong-confinement regime, consistent with the previous result that the ordered structure of strongly-confined water near the substrate accompanies the enhanced $\tau_r$[9]. Notably, the bcc DDAA begins to emerge in the same region ($d \sim$ 2 nm). Therefore, the bcc DDAA structure contributes to the solidity of the confined water. In other words, the solid-like behavior of strongly confined water results from the predominant presence of the bcc DDAA structure of ice-VII. Meanwhile, the increase of $\tau_r$ above 2 nm can be attributed to the change of the HB structure at the air/water interface as the elongation of the nano-meniscus progresses further[19].

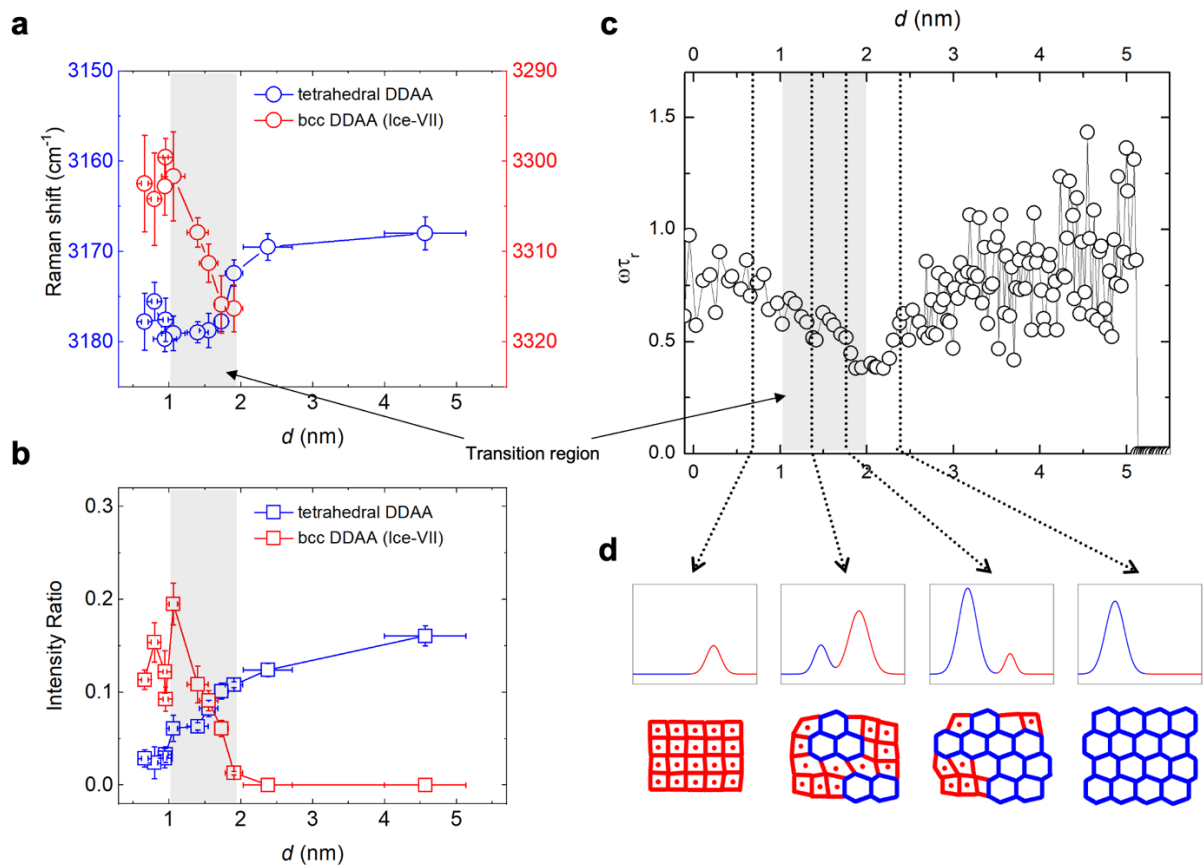

**Fig. 4. Confinement-dependent structural transition between two DDAA structures. a**, The Raman shifts versus $d$ show structural transition between tetrahedral DDAA (ice-$I_h$) and bcc DDAA (ice-VII). The grey area represents the transition region where the two DDAA structures interchange. **b**, The spectral area versus $d$ for both tetrahedral DDAA and bcc DDAA, which are normalized since the volume of confined water remains constant during experiments. **c**, Dimensionless mechanical relaxation time of confined water versus $d$ observed during tip retraction. **d**, The fitted curves for the two DDAA Raman signals (red; bcc DDAA, blue; tetrahedral DDAA) at specific values of $d$, and the corresponding schematics of molecular structure.

The ice formation observed in confined space is truly unique, compared to previous studies on confined water mostly focused on fixed confinement conditions using two-dimensional materials or nanopores. By controlling the confinement with sub-nanometer resolution, we directly observe the effect of confinement on water. Algara-Siller *et al.*[37] report that confined water between two sheets of graphene at room temperature formed of square ice. Although the experimental conditions are similar to ours, the structure of confined water is quite different from that of ice-VII. Kapil *et al.*[7] and Lin *et al.*[38] calculated the temperature-pressure phase diagram of monolayer water using first-principle simulation. There is square ice in their diagram, but not ice-VII. This is because bulky ice structures such as ice-VII cannot be formed in monolayer water. For water that is not restricted to a monolayer, Jiang *et al.*[11] performed simulation showing that confined water more than monolayer thickness can form ice-VII rather than square ice. Our results also consistently show that the it is consistent that the intensity of the ice-VII (bcc DDAA, red line near 3316 cm$^{-1}$ in Fig. 2a) is decreased in the strongest confinement region of $d = 0.66$ nm, since there is no space for ice-VII to exist in this region, the remaining signal may be attributed to square ice.

Furthermore, our results give insight into understanding the strange characteristics of one-dimensional confinement. The ice-like structure of confined water inside CNTs with a diameter of 3 nm or less was predicted[39] and observed[40] at room temperature. However, the confined water inside cylindrical nanopores such as Mobil Composition of Matter No. 41 (MCM-41)[41] or anodic aluminum oxide (AAO)[42] did not exhibit an ice phase but remained supercooled water even under the same degree of confinement as CNTs. Our finding suggests that an ice structure forms in strong-confinement, while supercooled water is favored in weak-confinement. Therefore, one-dimensional confinement with carbon walls acts as strong-confinement, whereas confinement with oxide walls acts as weak-confinement, potentially due to the wall characteristics such as wettability.

For instance, the interaction between the hydrophobic wall and water molecule is relatively weaker than the interaction between the hydrophilic wall and water molecule, so the intermolecular interaction between water molecules is dominant within hydrophobic confinement, resulting in a depletion layer. This hydrophobic force[43] may cause the strong-confinement-like effect on water under hydrophobic confinement.

We finally address the confinement effect. Ice-VII is a well-known high-pressure ice phase at room temperature, where the minimum pressure required for bulk ice-VII formation is 2.02 GPa[44]. We estimate the pressures exerted on the confined water and compare them with the pressure for stable bulk ice-VII as shown in Fig. 5a. The pressure is calculated from the measured shear-direction stiffness (Fig. S5) using the normal- and shear-hydration force relations through the Poisson ratio[45]. The stiffness of the confined water indicates that the confined water does not squeeze out when the distance between the tip and the mica substrate is below the capillary-condensation point (~ 2.7 nm). Thus, the confined water starts to pressurize. While the Laplace pressure dominates in the weak-confinement regime, while the hydration pressure dominates in the strong-confinement regime, As shown in Fig. 5a. Comparing the Raman shifts of bcc DDAA with those of bulk ice-VII (Fig. 2e), we immediately find that the pressure exerted on the confined water is only about 0.19 GPa when bcc DDAA network emerged in the confined water (i.e., $d \approx$ 2 nm). Notice that the minimum pressure required to form the bulk ice-VII is 2.02 GPa. Thus, the threshold pressure for the formation of bcc DDAA in strongly-confined water is about 10 times lower than that of the bulk ice-VII due to the confinement effect. Consequently, we establish a phase diagram of water under confinement as described in Fig. 5b. Note that the confined water directly transforms into ice-VII even though it passes the stable condition of ice-VI depicted as a darker grey region near 1 GPa in Fig. 5b. Similar direct ice transformation to ice-VII from the supercompressed water is reported[46] and the authors found that the local order of supercompressed water is similar to ice-VII than ice-VI. The shift of the phase boundary of ice-VII is associated with the substantial reduction in entropy of the molecules within the nano-confined geometry, resulting in the limited degrees of freedom and the structures of the HB network[47]. Recent *ab initio* MD simulation[7] reported similar results to ours, where the liquid-solid phase boundary for monolayer water (blue-dotted line in Fig. 5b) was shifted relative to that of bulk water due to confinement. Notice that the electrofreezing effect induced by the plasmonic hot-carriers or the static charges are not responsible for ice-VII formation, as they are not strong enough to cause crystallization (see Supplementary discussion S8).

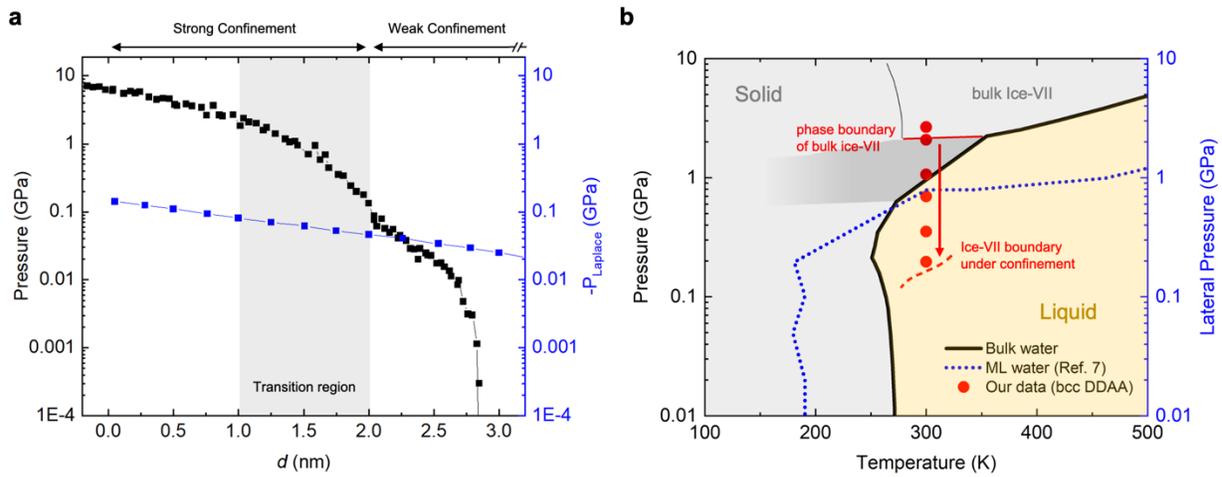

**Fig. 5. The confinement-induced shift of liquid-solid phase boundary for ice-VII and the applied pressure under confinement. a,** The estimated normal pressure (black) obtained from measured shear stiffness of confined water (Fig. S5) and the calculated negative Laplace pressure (blue) versus *d*. The grey region represents the transition region where the two DDAA signals are exchanged. The confinement below 2 nm distance is defined as the strong-confinement where the bcc DDAA emerges. **b,** The thick black-solid line is the liquid-solid phase boundary of bulk water. The blue-dotted line shows the simulated liquid-solid phase boundary for monolayer water[7]. The red-dashed segment represents the local phase boundary of ice-VII

expected from our data (red dots) under weak-confinement at 0.19 GPa pressure, which is shifted down in pressure from the bulk-Ice VII phase boundary (red-solid segment).

In summary, we have used TERS to address the structures of the HB network of confined water at room temperature. We have shown that the bcc DDAA, the unit structure of ice-VII, is formed under strong-confinement ($d < 2$ nm), while the tetrahedral DDAA, the unit structure of ice-I$_h$, dominates under weak-confinement. Notably, the bcc and tetrahedral DDAA structures undergo a structural transition as varying the confinement height $d$ with the transition width of 1 nm. The spatial distribution of HB networks (Fig. 3j) shows that ice-VII forms within the interior water rather than at interfaces of confined water. Furthermore, we have found that the pressure threshold for ice-VII formation (bcc DDAA) is 10 times lower than that of bulk ice-VII. Finally, our study on the structural transition of the HB network yields a new phase diagram of confined water, as shown in Fig. 5b. Our quantitative findings contribute to understanding the confinement-dependent changes in the complex HB networks in water, which would elucidate the origin of anomalous behaviors in low-dimensional water.

# Supplementary Information

1. **Materials and Methods**
2. **Supplementary Discussion**
    - **S1. RH dependence of approach curve**
    - **S2. Tip dependence in bcc-structure observation**
    - **S3. Determination of zero-point distance**
    - **S4. Bulk ice-phase references**
    - **S5. OH-stretching bands in full spectrum**
    - **S6. Spatio-spectral behaviors of DDA, DA and DAA**
    - **S7. Spatial distribution analysis process of HB network within nano meniscus**
    - **S8. Electrofreezing charge effect**
3. **Supplementary Figures**
4. **Supplementary References**

**Materials and Methods**

**TERS setup.** Figures 1a, b show the schematic of TERS setup with bottom-illumination configuration. TERS is realized by combining the home-made QTF-based shear-force AM-AFM and confocal Raman Spectroscopy. For shear-force feedback of *d* control, QTF is electrically excited by a function generator (33120A, Agilent), and its amplitude-phase response is monitored by a lock-in amplifier (7265, EG&G instruments). A mica substrate is mounted on the sample holder with a piezoelectric transducer attached for scanning. During scanning, the staying time between each data point is set long (~ 250 ms) because of a high quality factor (~ $10^4$) of the QTF.

The entire TERS setup is enclosed in the airtight humidity-controlled chamber. For Raman spectroscopy, a 532 nm laser beam is used for excitation, which is expanded to 12 mm diameter using two lenses after collimation. To reduce the unwanted background signals, we use only the large NA (> 1) component of the laser beam by blocking its small NA (< 1) part with an optical mask. High NA objectives are employed for high excitation/detection efficiency (PlanApo, 60x, 1.45NA, Olympus). The backscattered Raman signal is collected by the same objective, passed through a cut-off filter to block the Rayleigh scattering signal and detected by a fiber-coupled spectrometer (UHTS 300 VIS, 600lines/mm, Witec) with an EMCCD camera (DU970P, Andor).

**Tip fabrication** We have prepared the TERS tips in three different ways. First, the tips used in our main experiments are fabricated by e-beam evaporative coating of chromium (Cr) and silver (Ag), showing the EF above $10^5$. We select the cantilever (ATEC-CONT, Nanosensors) having an aspect ratio higher than the standard cantilever to obtain higher EFs. The cantilever is placed in a UV Ozone cleaner (UVC-150, Omniscience) for 30 min to remove organic contaminations from the surface before coating. Without the UV Ozone treatment, a strong signal from contaminants is observed during measurements, while the contaminants also affected the Ag-coating adhesion strength to the surface. The tips are coated using an E-beam evaporator (KVE-E4006L, Korea vacuum tech) at $10^{-6}$ Torr. 5 nm thick Cr is first coated as an adhesive layer, and then 30 nm thick Ag is coated on the Cr layer. The average tip radius of coated cantilever is about 37 nm. The coated tip is glued with UV adhesive (NOA88, Norland) to one prong of the QTF immediately after coating. The experiment is conducted within 3 hours after tip fabrication to minimize oxidation of silver surface which caused reduction of EF.

Second, etching of the tip is done using 0.1 mm diameter Ag wire (99.99%, Goodfellow). 0.1 mm diameter Au wire (99.9%, Sigma Aldrich) is made in a circular form with a diameter of 1 cm and used as a counter electrode. A 1:4 mixture of perchloric acid (70%, Sigma Aldrich) and ethanol (99.9%, Daejung) is used as an etchant. A 6 V of dc voltage is applied between the Ag wire and counter electrode, after they are immersed in the etchant (0.5 mm and 0.2 mm from the surface, respectively). While monitoring the etching current, terminating the etching process is done when the etching current reaches 0.15 mA. The fabricated tip is washed with deionized water (Arioso Power I, Human corporation). Third, sputter coating (spt-20, Coxem) is conducted at the pressure of $10^{-2}$ Torr. After 30 min of UV Ozone cleaning, 35 nm thick Ag is sputtered on the cantilever at 6 mA current flow.

**Sample preparation.** The cleaved muscovite mica (SPI supply) used as a hydrophilic surface is attached to the cover glass with the optically transparent glue (NOA88, Norland). It is known that there is a charge imbalance on the freshly cleaved mica surface. This charge effect is negligible in our experiment because we discharged the surface by stabilizing the system in a closed chamber in ambient conditions for at least 3 hours before experiments. The scanning image of the mica sample obtained by a commercial AFM (NX-10, Park Systems) is shown in Fig. S6.

**Data acquisition.** QTF is calibrated using the energy-balance methods[48] ($\alpha$ = 0.66 Å/mV, amplitude = 0.29 nm at $\omega/2\pi$ = 32,745 Hz). The chamber is purged with dehumidified nitrogen gas (RH < 1 ppm) before experiments. Deionized water is used to set the relative humidity by controlling the flow rate of two gases, pure nitrogen gas and water-saturated gas. Raman measurements are conducted with 50 µW laser power to minimize photo-induced reactions. Relatively high power (5 mW) is only used to measure the bulk-water Raman signal. Each spectrum is measured for 3 min while the distance $d$ is kept by AFM-PID control. Raman signals are baseline-corrected by the improved asymmetric least squares (IAsLS) method. OH stretching band signal is fitted with 7 Gaussian curves with Origin Pro. Two fitted curves of free-OH signal and $CH_n$ signal is omitted in main figure by subtracting their Gaussian curves from raw data.

**FDTD simulation.** We use the finite-difference time-domain (FDTD) method to calculate the 'hotspot' size near the TERS probe using the COMSOL Multiphysics. The probe model has a cone-shaped Si core of a 1:4 aspect ratio with a tip radius of 10 nm, and 30 nm thick Ag coating is applied on the outside. We set the fused quartz as a substrate and performed the simulation while varying $d$ from 0.05 nm to 4.5 nm. As a background initial wave, the plane wave at 532 nm wavelength with its wave vector parallel to the substrate is employed. The refractive index of Si[49], Ag[50], and fused quartz[51] are referred from each reference.

**SERS experiment.** 1000 $\mu L$ of 100 nm AgNPs (Sigma Aldrich) solution is centrifuged at ~ 15,000 rpm for 10 min, discarded 950 $\mu L$ of remaining solution from top, redispersed with 950 $\mu L$ distilled water, and centrifuged again. The AgNPs solution is sufficiently purified at least 5 times to detect nano-size captured water between NPs[12]. To measure Raman spectrum, a small droplet of purified AgNPs solution (~ 2 $\mu L$) is deposited on the silicon wafer substrate, the sample is loaded in the temperature-controlled chamber (HCS302, INSTEC). After sufficiently dried in ambient condition, Raman measurement is conducted with 50 $\mu W$ laser power and 90 second integration time.

**Supplementary Discussion**

## S1. RH dependence of approach curve

We confirm systematically the effect of coalescence caused by a thin film of water on the surface. We assume the pre-adsorbed water film on mica is a monolayer thick (~ 0.3 nm) as is confirmed in the previous experiment[36]. However, it has been also known that the thickness of water film on silver is thicker than that on mica[52]. The thick pre-adsorbed water film causes two problems in precise distance control. First, if the size of the nano-meniscus is larger than the 'hotspot' size, the interpretation of the Raman signal would become abstruse because the signal would come only from a part of the meniscus. Second, QTF could be easily saturated, due to its high Q-factor nature, even by a small force from the thick pre-adsorbed water film. In order to avoid such problems, we had to create a nano-meniscus as small as possible by reducing the relative humidity (RH).

The typical approach curves obtained at different RH, the measured critical distance $d_c$ where the tip-oscillation amplitude drop occurs, and the corresponding amplitude magnitude at $d_c$ are shown in Fig. S7. All the amplitudes are normalized by their free oscillation amplitude. At high RH, all parameters such as the amplitude drop, the measured $d_c$, and the rupture distance also increased, obviously indicating the increase of the thickness of pre-adsorbed water film. Constructing a quartz-mica (tip-mica) system[36] for comparison, the results are very different because the measured $d_c$ and the rupture distance are approximately 1.2 nm and 2.2 nm, respectively at 10% RH. This difference shows that the pre-adsorbed water film developed on the silver surface is thicker than that on mica.

Additionally, the calculated $d_c$ (< 0.6 nm) from the Kelvin equation at 1 ~ 10 % RH is much shorter than the measured $d_c$ (see Fig. S7d). This is because the nucleation effect and the coalescence effect are combined in the measured $d_c$, but even with reference to the previously reported thickness of water film (~ 1 nm) on silver[52] at given RH, there is still a large discrepancy between the calculated $d_c$ and the measured one above 5 % RH. There are several possible, but not clarified, causes such as the uncertainty of zero-point determination, the laser trapping effect and the interfacial tension effect, that may contribute to such discrepancy. In our experiments, we conduct all the experiments at 1.2 ± 0.5 % RH because (i) it is confirmed there is no significant change in the measured $d_c$ below 2 % RH, and (ii) the volume of the meniscus calculated by the Young-Laplace equation at 2 % RH does not exceed the 'hotspot' size (See Fig. S1).

## S2. Tip dependence in bcc-structure observation

We have conducted the same experiment to check the tip dependence in Raman observation of the bcc structure using the tips fabricated by different methods; sputter coating and etching. The results are shown in Fig. S8. Similar to the main figures in the text, the $CH_n$ stretching signal and the free-OH signal are omitted. Although a similar OH stretching band signal is obtained using the sputtered tip, there is one important difference. In particular, the bcc DDAA signal is not obtained reproducibly with sputtered tip. The reason is that the proper confinement conditions may not established between tip and mica. As shown in the inset of Fig. S8a, the surface roughness is quite large and there is a relatively large gap which is invariant with confinement between the silver metal chuck. Notice that the peak at 3200 cm$^{-1}$ is not changed with confinement, showing that proper confinement is not established by the sputtered tip. On the other hand, the bcc DDAA signal is measured using the etched tip. As can be seen from the SEM image, this is considered as due to the surface smoothness of the etched tip. Although the exact peak positions are not same as those obtained by the e-beam evaporation tip, it showed similar behaviors, implying that the HB network of confined water is highly dependent on the confinement environment, as is known[6,33,53].

## S3. Determination of zero-point distance

The zero-point ($d = 0$) is defined with reference to the contact point or the point of maximum energy dissipation. The energy dissipation, the integration of the non-conservative force over one period, is a characteristic mechanical process in the dynamic AFM mode operation. The energy dissipation per cycle is given by[18],

$$\frac{E_{\text{dis}}}{kA_0^2} = -\frac{\pi}{Q}\left[\frac{A}{A_0}\cos\theta - \frac{\omega}{\omega_0}\left(\frac{A}{A_0}\right)^2\right]$$

where $A$ and $\theta$ are the amplitude and phase of the probe, respectively, $A_0$ is the peak amplitude at resonance, $\omega_0$ the resonance angular frequency, $\omega$ the operating angular frequency, $Q$ the quality factor of the probe, and $k$ the stiffness of the probe. For our QTF, $A_0 \sim 0.29$ nm at $\omega \sim 32,745$ Hz, $Q \sim 9000$, $k \sim 43000$ N/m. The typical data of the approach curve and the corresponding dissipation energy are shown in Fig. S9. Though the absolute standard for measurement of a contact point in AFM still lacks, we have defined the contact point as the position of maximum energy dissipation (or maximum non-conservative interaction), which is one of the important criteria to identify the surface properties using AFM.

Then, because there is a tightly bound monolayer thickness of water on each hydrophilic surface, on the mica surface and on the silver tip surface, the zero point $d_{\text{zero}}$ can be defined as two water monolayer thickness away from the contact point $d_{\text{contact}}$, that is $d_{\text{zero}} = d_{\text{contact}} - 0.6$ nm although there can be slightly different definitions of the zero-point[36,54]. In addition, there are inevitable experimental uncertainties in determining the zero-point such as the elasticity of the materials and the meaning of 'contact' at the molecular scale. Nonetheless, by comparison to the previous studies[36,54], we estimated there can be at most a maximum error of $\pm 0.5$ nm. Our interpretation of the simultaneously measured Raman signal of the nano-meniscus and the relaxation time is not altered regardless of how the zero-point is defined within experimental error.

## S4. Bulk ice-phase references

The DDAA signals of each ice phase in Fig. 2e are measured under the following conditions. Ice I[23,55] obtained at 0.1 MPa to 1 GPa and at 70 to 246 K. Ice II[23,56] at 0.28 GPa and at 225 K. Ice III[23,56,57] at 0.28 GPa and at 246 K. Ice IV[29] at 0.81 GPa and at 80 to 140 K. Ice V[23,56] at 0.42 GPa and at 246 K. Ice VI[23] at 0.62 GPa and at 246 K. Ice VII[44] at 2.04 to 30 GPa and at room temperature. Supercompressed ice-VII[46] at 1.72 GPa and at room temperature. Ice VIII[22,31] at 2.8 to 18.9 GPa and at 80 K. Ice IX[57] at 0.28 GPa and at 155 K. Ice XI[58] at 0.1 MPa and at 60K. Ice XII[59] at 1 kPa and at 80 to 110 K. Ice XIII[60] at 1 kPa and at 80 K. Ice XIV[59] at 1 kPa and at 80 to 110 K. Ice XV[61] at 0.9 GPa and at 70K.

## S5. OH stretching bands in full spectrum

The full spectrum contains various signals other than confined water, but we omitted these signals in the main figures for clarity. In the low-frequency region of 100 – 450 cm$^{-1}$, the Raman signals of mica are at 195 cm$^{-1}$, 265 cm$^{-1}$, and the lattice vibration mode of confined water is at 210 cm$^{-1}$ as shown in Fig. S10. Mica signals are easily distinguished from the lattice vibration mode of confined water since the mica signals are not changing and are observed before ($d$ = 7 nm) the nano meniscus is formed. We have subtracted the mica signals at $d$ = 7 nm from every raw data and have plotted in Fig. 2a. In the high-frequency region of 2800 - 3600 cm$^{-1}$, the impurity signal is at 2900 cm$^{-1}$, the OH stretching band of confined water is at 3000 - 3500 cm$^{-1}$, and the free OH mode of mica is at 3610 cm$^{-1}$. The impurity signal is an aliphatic CH stretching mode that changes when using a tip with high enhancement factors as shown in Fig S10a. This is not $d$-dependent, but rather time-dependent as demonstrated in the next paragraph. The possible sources of impurities are atmospheric carbon contaminant (ACC) or carbonaceous decomposition products, but their exact origin and specific molecular structure are still unknown[25–27,62–64]. We have subtracted each impurity signal at $d$ from the corresponding raw data and have plotted it in Fig. 2a. Meanwhile, the free OH mode of confined water is buried by the same signal from mica. Since most of the free OH mode came from mica, we have subtracted the free OH mode at $d = 7$ nm from every raw data and have plotted it in Fig. 2a.

The intensity of the $CH_n$ stretching band is decreasing over time as shown in Fig. S10. This is a photo-induced chemical transformation, meaning that the aliphatic CH bond at 2900 $cm^{-1}$ is decomposed or transformed to the other bond[65]. We measure the Raman signal of the tip placed far enough from the substrate (> 50 μm) while irradiating laser continuously on purpose to check the effect of the photo-induced chemical transformation itself or its residue on the Raman signal of water, especially on the OH stretching band. To accelerate the chemical transformation process, the laser power is used at 200 μW, higher than the main experiment, and the integration time of each spectrum is 1 min. As shown in Fig. S11, the aliphatic CH mode at 2900 $cm^{-1}$ is decreased over time. Note that there is no additional peaks are generated in the region above 3000 $cm^{-1}$ and the background signal does not change as well, meaning that the chemical transformation residues of the aliphatic CH bond do not disturb to measuring of the OH stretching band of water and the plasmonic structure does not change during laser radiation.

## S6. Spatio-spectral behaviors of DDA, DA and DAA

The Raman shift and the spectral area of DAA and DDA signals are shown in Fig. S12 a, b. The spectral area of both DAA and DDA signals show an abrupt decrease near 2 nm. According to our spatial distribution analysis of HB network (Fig. 3), we attribute the decrease of spectral area to the air/water interfacial volume that change with *d*. It is interesting to note that the intensity of DAA is comparable to other kinds of water, despite it being reported that a portion of the DAA water is very low in many systems including air/water interface[13]. From the result of the hydrogen-bond measurement at the flat air/water interface obtained by the sum-frequency generation spectroscopy, it is confirmed that DAA water is merely exists at the air/water interface. But it has been also reported that the DAA water can be abundant in a high curvature environment[66].

The Raman frequency and the spectral area of the DA signal with respect to *d* are shown in Fig. S12c, d. As the confinement gets stronger, the DA water binds more strongly to the surface while the portion of DA signal in the meniscus increases. The increase in the portion of the DA signal can be seen as an increase in the solid/water interfacial volume that occurs as *d* decreases. In particular, the intensity of DAA and DDA water starts to decrease at 2 nm while the intensity of DA water starts to increase at 2 nm, which indicates that the DAA and DDA water at the air/water interface close to the tip and the mica change to the DA water at the solid/water interface through the hydrogen-bond rearrangement. It is consistent with this interpretation that the increasing amount of DA water and the decreasing amount of DAA and DDA water are almost the same.

## S7. Spatial distribution analysis process of HB network within nano meniscus

We have speculated the spatial distribution of four types of water (DDAA, DDA, DAA, and DA) within the meniscus by comparing the enhancement factor (EF)-weighted volume of meniscus and Raman signals. The Raman intensity has the following relationship, $I_{\text{Raman}} = I_0 \sigma N \Omega E_{\text{EF}}(\vec{r}) D(\lambda) G(\lambda)$, where $I_0$ is the far-field intensity, $\sigma$ the Raman cross-section, $N$ the number of molecules, $\Omega$ the solid angle, $E_{\text{EF}}(\vec{r})$ the position-dependent enhancement factor, $D(\lambda)$ the detector efficiency, and $G(\lambda)$ the geometrical effect. Since $I_0, \sigma, \Omega, D(\lambda), G(\lambda)$ are under the same condition for four types of water, with constant density assumption, the Raman signal is proportional to the EF-weighted volume. The spatial distribution of EF was calculated by FDTD simulation (COMSOL), and the meniscus profile was calculated by solving the Young-Laplace equation[20].

The inside the meniscus is divided into three regions. Air/water (a/w) interfacial water region, solid/water (s/w) interfacial water region, and interior water region. The thickness of the a/w interfacial water $t_{a/w}$ and the thickness of the s/w interfacial water $t_{s/w}$ is set as free parameters. At given $t_{a/w}$ and $t_{s/w}$, the volume of the three regions (a/w interfacial water, s/w interfacial water, and interior water) is determined. for example, the green region (air/water interfacial water region) in Fig. 3d-f is determined by $t_{a/w}$, and the blue region (solid/water interfacial water region) in Fig. 3d-f is determined by $t_{s/w}$, and the remaining region is the interior water region.

Then, the EF-weighted volume map is obtained by integrating the infinitesimal volume multiplied by the EF map (Fig. 3a-c) over each partial region, a/w, s/w, or interior regions (Figs. 3d-f). Since the EF-weighted volume is proportional to the Raman intensity, the normalized EF-weighted volume and the normalized Raman spectrum of confined water are compared to assign which Raman signal to which spatial region. the spatial region inside the nano meniscus and the Raman signal of the nano meniscus cannot be matched in a one-to-one correspondence since the former is 3 and the latter is 5. Therefore, it is assumed that the DDA signal and DAA signal are attributed to the same spatial region and matched with the EF-weighted volume of the air/water interfacial water region[66]. The DA signal is matched with the EF-weighted volume of the solid/water interfacial water region[12]. And, it is assumed that the two DDAA signals are attributed to the same region and matched with the EF-weighted volume of the interior water region[33,35].

The error map can be calculated by subtracting the normalized EF-weighted volume of one of the spatial regions and the corresponding normalized Raman intensity of the confined water signal. then there are three error maps for each spatial region and the total error map is the sum of the absolute value of three error maps. The total error maps for $d = 4.5$ nm, $d = 2.5$ nm, and $d = 1.0$ nm are shown in Fig. 3g-i, respectively. The error value is not defined in the dark purple region at the lower part of Fig. 3i because $t_{a/w}$ or $t_{s/w}$ are too thick and the sum of the partial volume exceeds the total volume of the meniscus.

At the purple point in the total error map (purple in Fig. 3g-i), the total error has its minimum value. These points along $d$ are shown in Fig. 3k, and the EF-weighted volumes and Raman intensity of confined water at these points with respect to $d$ are shown in Fig. 3j.

## S8. Electrofreezing charge effect

Solidification of water by the electric field is known as the electrofreezing phenomenon. The required electric field for electrofreezing of bulk water at room temperature is about 1 V/nm. Freshly cleaved mica has a very high surface potential of -10V or less due to the surface-charge imbalance. However, when cleaved mica is exposed to the air, the surface charge is neutralized within a few hours and the stabilized surface potential increases to -50 ~ -20 mV. In our experiment, the surface charge is neutralized by stabilizing for several hours before performing measurements. Therefore, the surface charge of mica is not considered as the cause of solidification because it is 40~100 times lower for electrofreezing to occur. Note that the photo-induced charge may produce electric field on the confined water. Under our experimental conditions (50 $\mu$W laser power with 2 $\mu$m diameter of beam spot), however, the electric field induced by the excitation laser is only $7.7 \times 10^4$ V/m. Even considering the field enhancement effect based on the FDTD simulation results, it is expected not to exceed $10^7$ V/m.

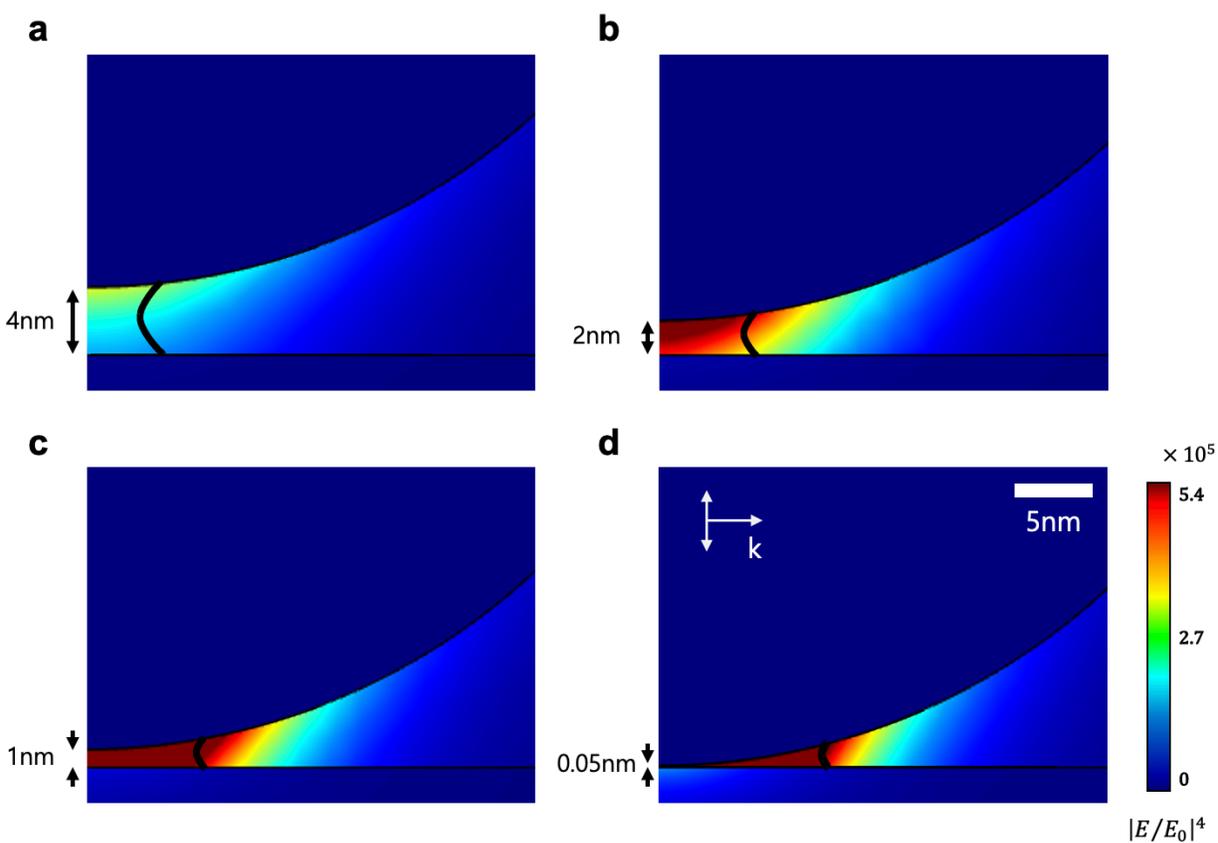

**Fig. S1.** FDTD simulation results of the electric field distribution at the apex of a silver coated tip and the respective nano-meniscus geometry (thick black solid line) calculated by the Young-Laplace equation. The tip-substrate distances are 4, 2, 1 and 0.05 nm, respectively, in **a-d**.

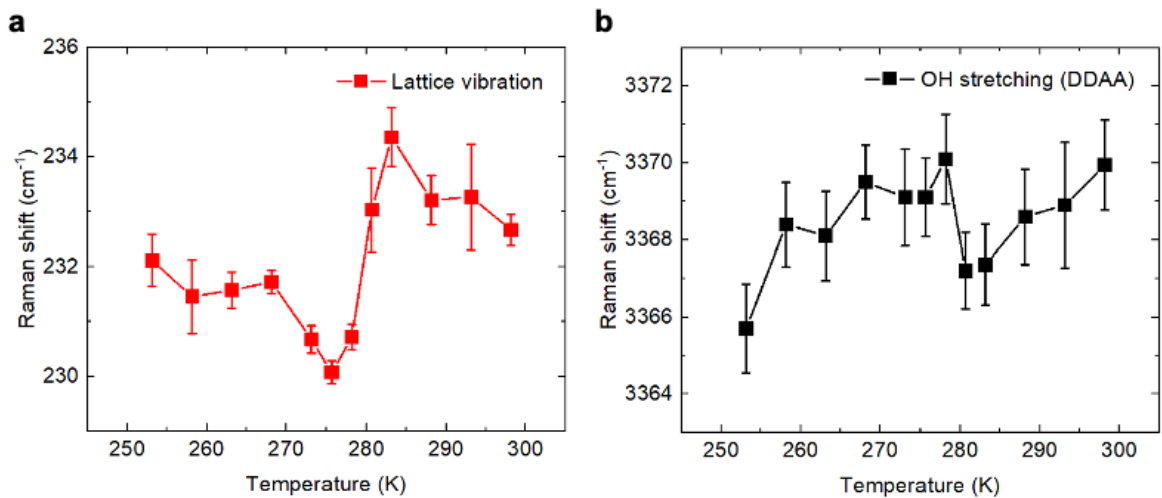

**Fig. S2. a.** Raman shift of the lattice vibration signal. **b**, DDAA peak at OH stretching mode according to temperature.

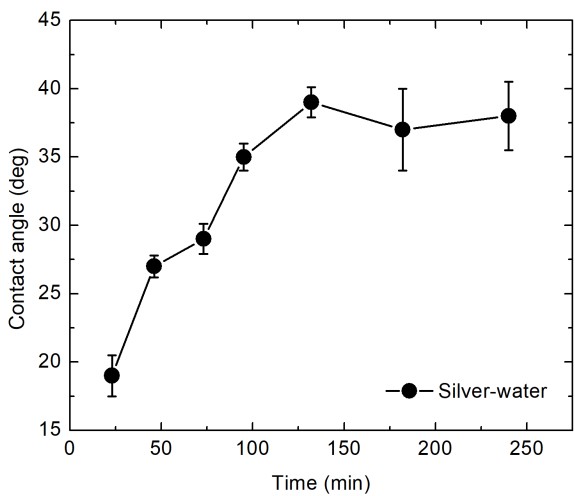

**Fig. S3.** Changing of Silver-water contact angle with time. Each data was measured three times and averaged.

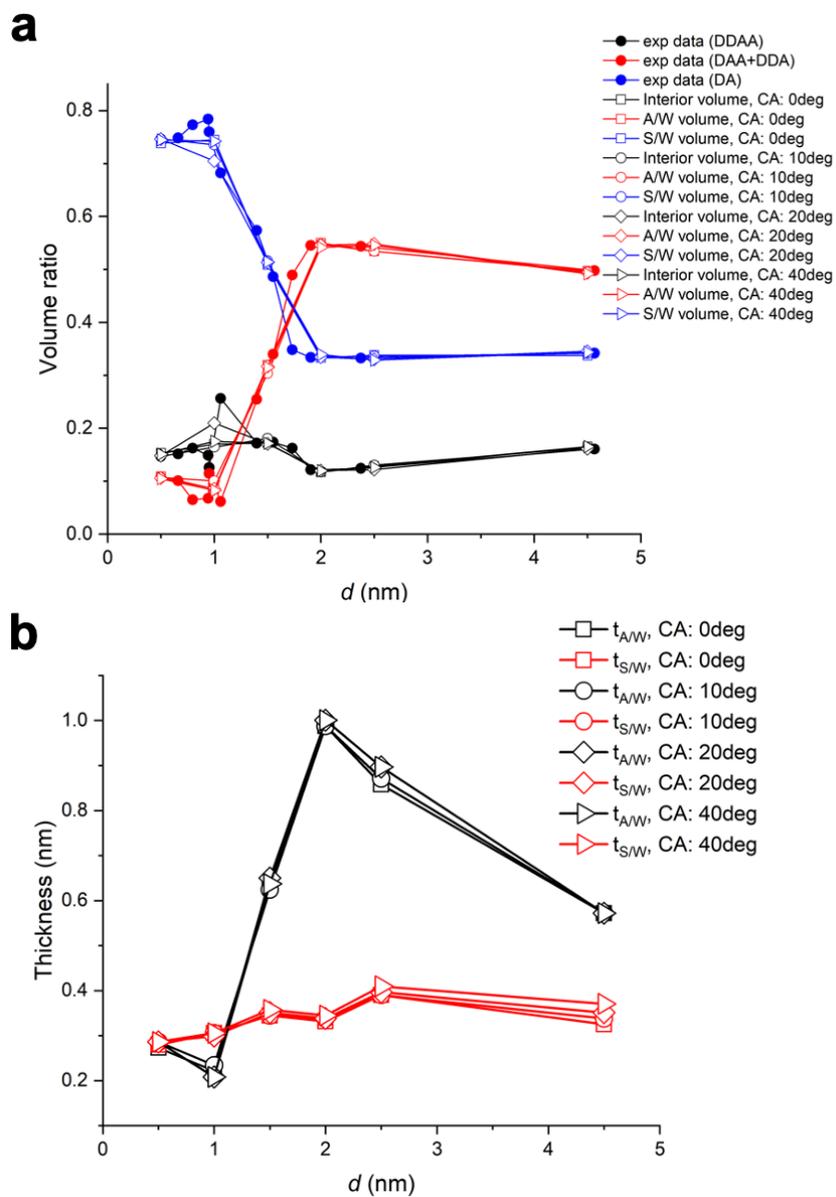

**Fig. S4. a**. Results of Spatial distribution analysis according to tip-water contact angle. mica-water contact angle was zero in all cases. **b**, Optimized parameters ($t_{a/w}$, $t_{s/w}$) according to tip-water contact angle.
35

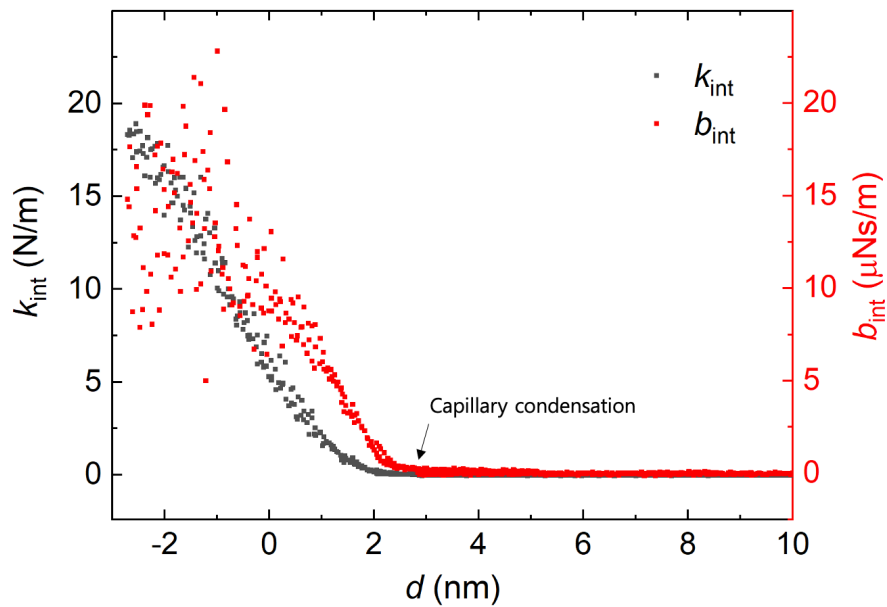

**Fig. S5.** Elastic stiffness coefficient $k_{int}$ (black) and damping coefficient $b_{int}$ (red) due to the meniscus that are obtained from the amplitude and phase signal of QTF in the shear-mode AFM operation. The arrow indicates the position of capillary condensation of the meniscus where the strong elastic and damping coefficient emerge with respect to bulk water.

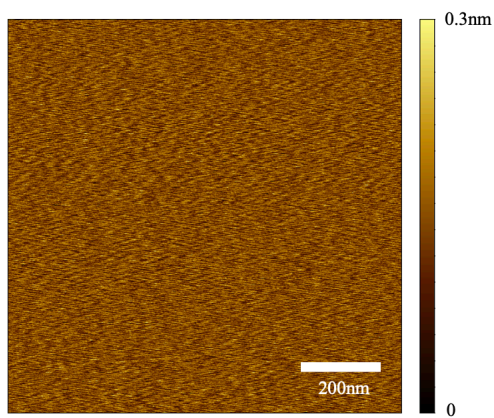

**Fig. S6.** Noncontact mode scanning image of mica obtained by a commercial AFM. The image shows the height uniformity of mica and the measured root-mean-square roughness is 0.41 Å.

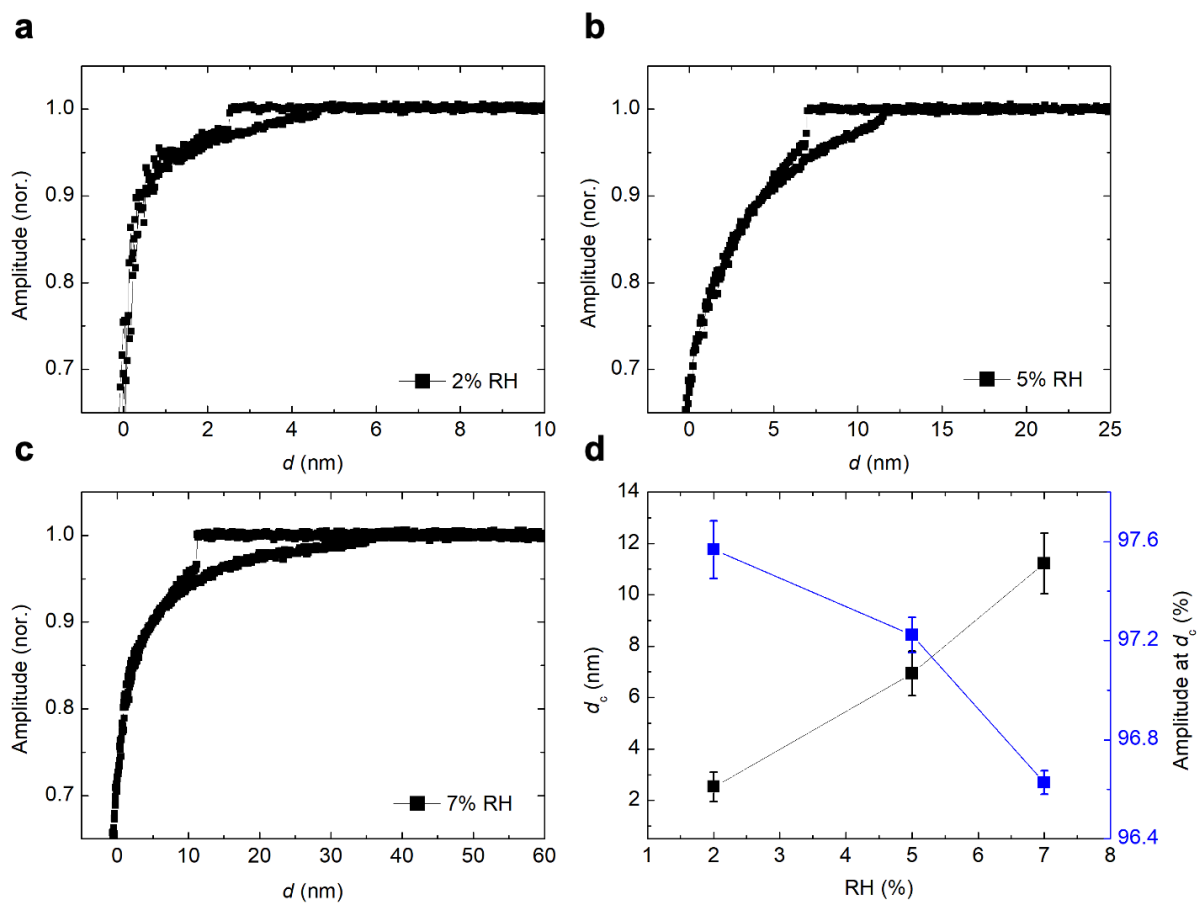

**Fig. S7. a**, **b**, **c.** Normalized approach curves at three different relative humidity (2%, 5%, and 7%, respectively). Both the critical distance at which water vapor condenses upon approach of the tip to the substrate and the rupture distance at which the water bridge ruptures upon retraction of the tip become longer as RH increases. **d**, The measured critical distance $d_c$ (i.e. distance where sudden drop of the tip-oscillation amplitude occurs) and the oscillation amplitude at $d_c$ (i.e. amplitude drop measured at $d_c$) against the relative humidity. The error bars are estimated from measurements with three different tips.

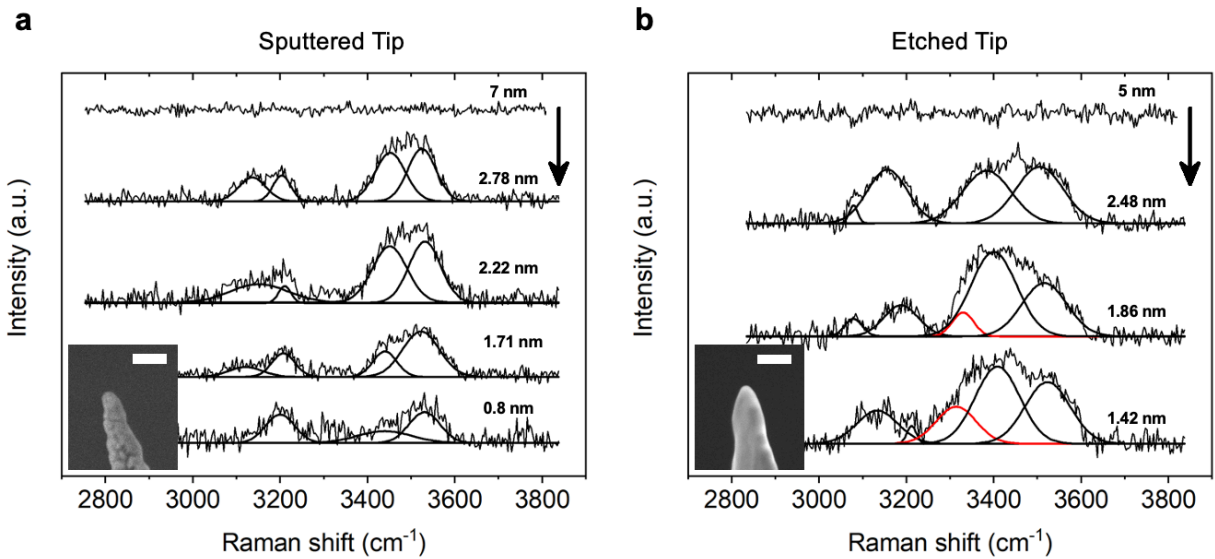

**Fig. S8.** Typical Raman signal of confined water obtained by sputtered tip and etched tip. The $CH_n$ stretching signal and free-OH signals are omitted for clear comparison. The arrow indicates the direction of tip approach. **a**, The signal obtained by the Ag-coated tip using sputter. **b**, The signal by the Ag-wire etched tip. Each inset shows the SEM image of the tip (scale bar is 200 nm).

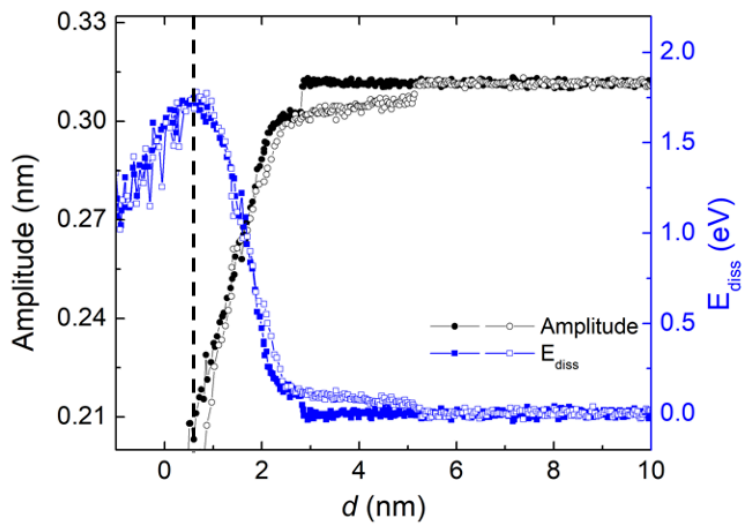

**Fig. S9.** Energy dissipation and tip-oscillation amplitude versus $d$. The contact-point is determined as the point of maximum energy dissipation, from which the zero-point ($d = 0$) is defined with reference to the contact-point as $d_{\text{zero}} = d_{\text{contact}} - 0.6$ nm.

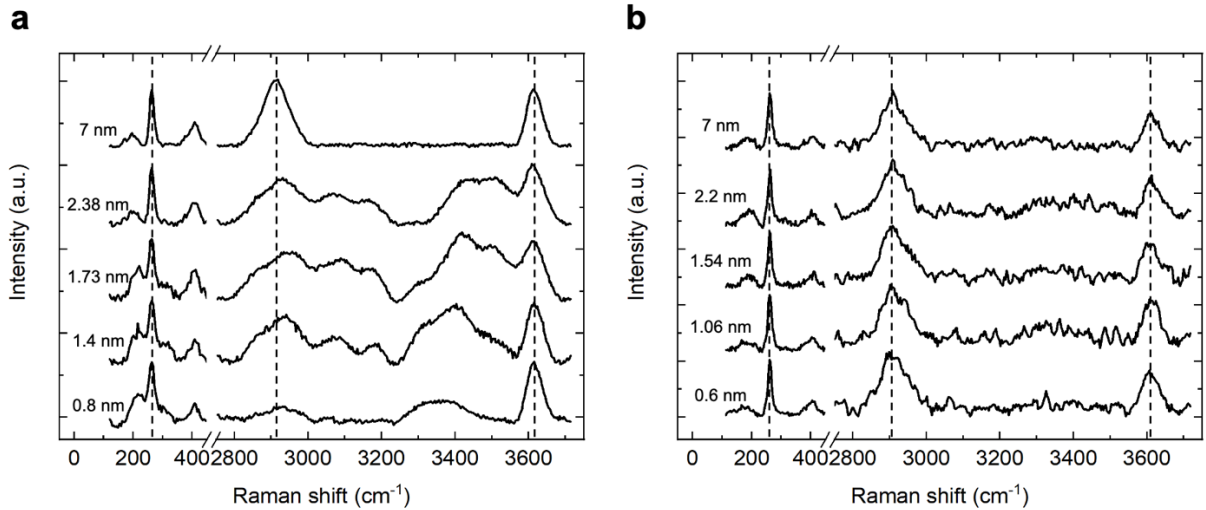

**Fig. S10.** Full spectrum of the nano meniscus in low-frequency region and high-frequency region. The mica signals at 195 cm$^{-1}$ and 265 cm$^{-1}$, and the lattice vibration mode at 210 cm$^{-1}$ are included in the low-frequency region. The OH stretching modes of confined water at 3000 – 3500 cm$^{-1}$, the CH$_n$ modes of impurities at 2900 cm$^{-1}$, and the free-OH mode of mica at 3610 cm$^{-1}$ are included in high-frequency region. **a**, The Raman signal obtained with the high (~$10^5$) EF tip. The peak position and the intensity of the CH$_n$ stretching band signal are changed due to the photo-induced chemical transformation. **b**, The signal with the low ($10^2 \sim 10^4$) EF tip. The positions of both CH$_n$ and free-OH signals are not changed versus $d$.

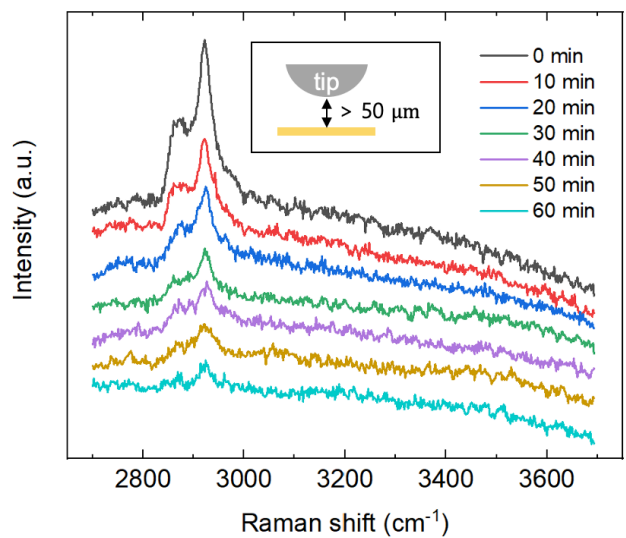

**Fig. S11.** Raman signal of the tip measured over time under continuous laser exposure.

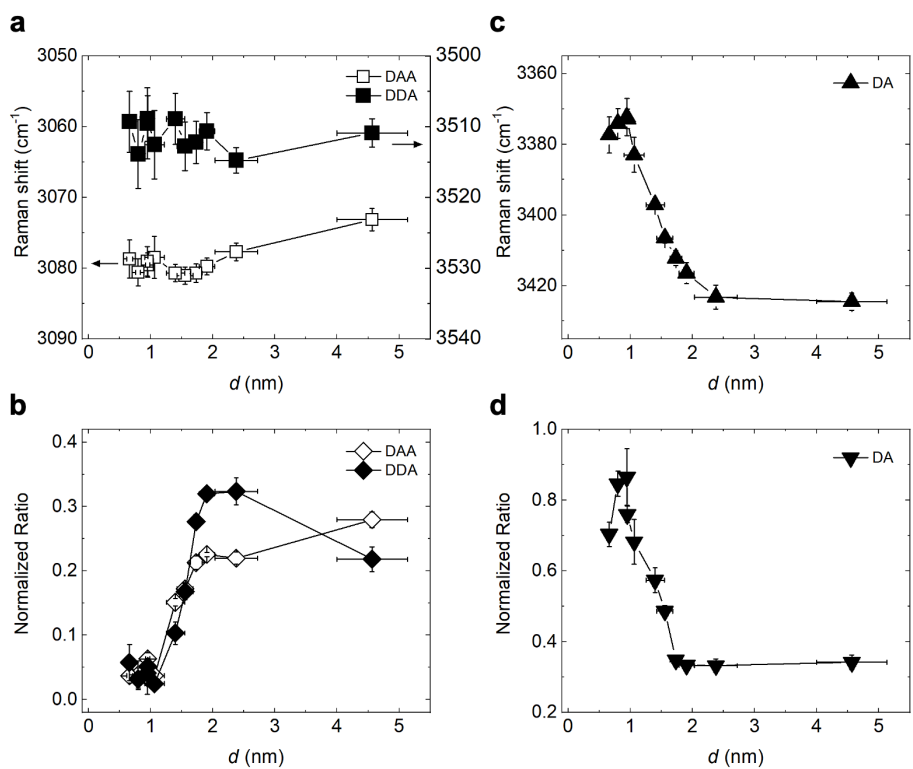

**Fig. S12.** Plots of the spectral peak shift versus $d$ of **a** DAA and DDA, and **c** DA. The normalized spectral area versus $d$ of **b** DAA and DDA, and **d** DA are plotted.

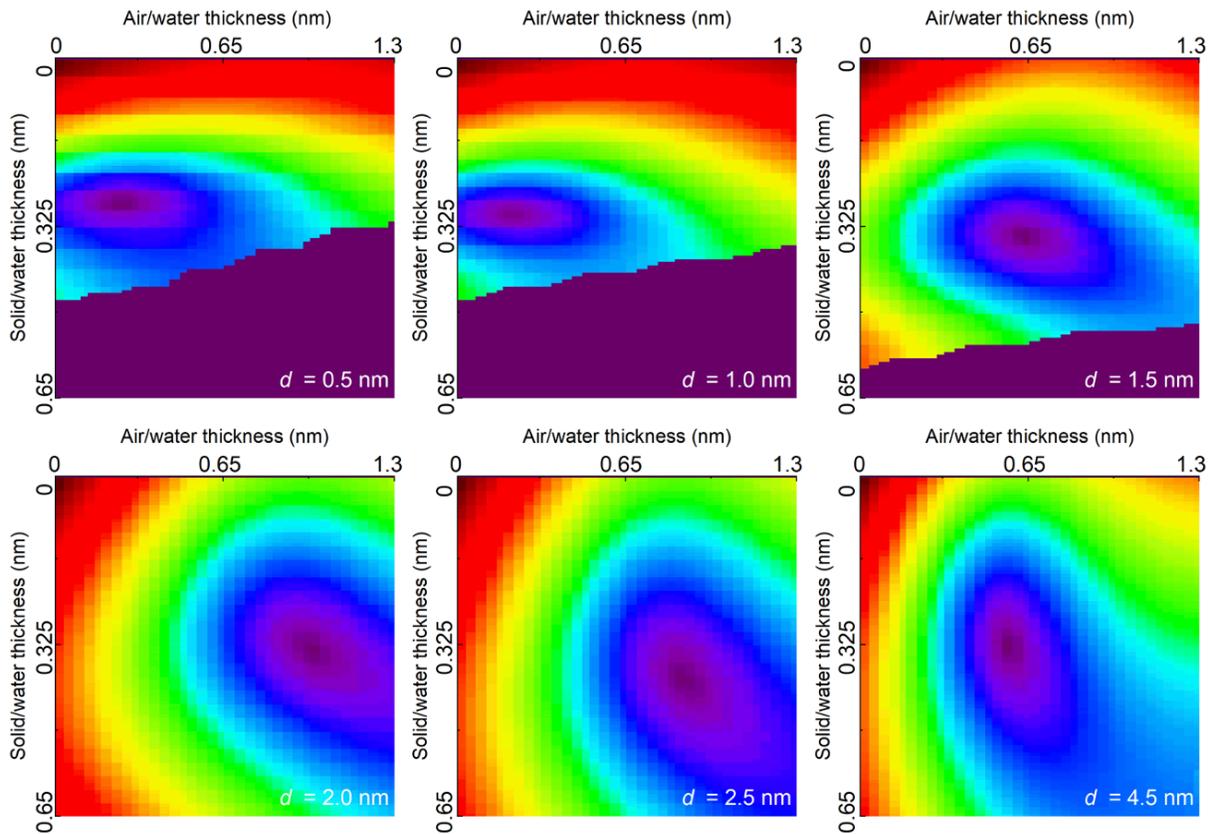

**Fig. S13.** Total error map of the spatial distribution analysis at each *d*. The error is defined as the sum of the difference between the experimental data and the calculated EF-weighted partial volume of meniscus. The error is not defined in the dark purple area at the lower part of figures because the sum of the partial EF-weighted volume exceeds the total EF-weighted volume. The error has its minimum value at violet area.

# Supplementary References

48. Dagdeviren, O. E., Miyahara, Y., Mascaro, A. & Grütter, P. Calibration of the oscillation amplitude of electrically excited scanning probe microscopy sensors. *Rev. Sci. Instrum.* **90**, 013703 (2019).

**Data availability** The data sets generated and/or analyzed during the current study are available from the corresponding authors on request.

**Acknowledgments** The authors acknowledge Dohyun Kim, Sangmin An, Sunghoon Hong and Qhwan Kim for helpful discussions. The SEM images of the tip were obtained at the Research Institute of Advanced Materials, Seoul National University. This work was supported by the National Research Foundation of Korea (NRF) grant funded by the Korean government (MSIP) (No. 2016R1A3B1908660). M.L. was suproted by the Overseas Dispatch Program of Chungbuk National University in 2022. W.J. appreciates the support by the School of Engineering and Applied Sciences of Harvard University during his visit.

**Author contributions** J.H. and W.J. designed and led the project and wrote the manuscript with M.L. and X.Z.. J.H. carried out experiments, and J.H, D.S., M.L., X.Z. and W.J. performed data analysis. All authors discussed and commented on the manuscript.

**Competing interests** The authors declare no competing interests.

**Additional information** Supplementary Information accompanies this paper and is available at the assigned site.